\documentclass{optica-article}

\journal{opticajournal} 

\articletype{Research Article}

\begin{document}

\title{Time-periodic (Floquet) systems in classical wave physics and engineering: Opinion}

\author{Francesco Monticone\authormark{1,*}, Dimitrios Sounas\authormark{2}, and Matteo Ciabattoni\authormark{1}}

\address{\authormark{1} School of Electrical and Computer Engineering, Cornell University, Ithaca, NY 14850, USA\\
\authormark{2} Department of Electrical and Computer Engineering, Wayne State University, Detroit, MI 48202, USA\\}

\email{\authormark{*}francesco.monticone@cornell.edu} 


\begin{abstract*} 
The study of classical waves in time-periodic systems is experiencing a resurgence of interest, motivated by their rich physics and the new engineering opportunities they enable, with several analogies to parallel efforts in other branches of physics, e.g., Floquet-engineered quantum materials and time crystals. Here, we first briefly review some of the most prominent features enabled by time-periodic modulations, and we then focus on two specific areas, namely, time-varying systems to break reciprocity and to overcome various theoretical limitations and performance bounds, discussing their current status, challenges, and opportunities.
\end{abstract*}

\section{Introduction}
 When a classical wave-physics system is modulated in time, what basic physical aspects are affected? The temporal modulation breaks continuous time-translation symmetry and, therefore, frequency, as well as energy and/or photon number, do not need to be conserved \cite{pendry2023photon,galiffi2022photonics}. In fact, in many cases, the system may no longer be considered passive, as the temporal modulation may pump energy into the system, as for example in parametric amplifiers. If the modulation is periodic with frequency $\Omega$, well-defined harmonics are generated with frequencies $\omega + n\Omega$ (where $ n \in \mathbb{Z}$). The field eigenmode of the time-periodic system, also known in the photonics literature as a ``photonic time crystal,'' can then be represented as a sum of these frequency harmonics and the dispersion diagram (band diagram) becomes periodic along the frequency axis, with the possibility of band-gaps opening along the wavevector/momentum axis when different dispersion branches cross, analogous to the frequency band-gaps of space-periodic systems. These bandgaps have been observed experimentally at microwave frequencies \cite{park2022revealing,reyes2015observation,wang2023metasurface} and many of their intriguing implications, including their topological properties, have been studied theoretically in recent years \cite{lustig2018topological,lustig2023photonic,lyubarov2022amplified,asgari2024theory}. 
 One can also look at this scenario from a coupled-mode-theory standpoint: if a certain mode is initially excited in a time-invariant system, the application of a time-periodic modulation can then induce coupling (or “interband photonic transitions”) between modes at frequencies separated by a multiple of $\Omega$ \cite{winn1999interband,dong2008inducing}. Transitions between a ladder of modes also enable the creation of a synthetic (frequency) dimension, whose implications to study high-dimensional physics in low-dimensional systems have been extensively explored \cite{yuan2018synthetic}. Moreover, these transitions can also occur between modes having opposite frequencies, for instance if $\omega -\Omega = -\omega$, i.e., $\omega = \Omega/2$ (corresponding to a momentum band-gap), which implies that the mode couples with itself, as positive and negative frequencies represent the same physical oscillation (due to the reality of classical fields), and energy can be extracted from (or transferred to) the temporal modulation \cite{Dovic1971generation,silveirinha2023hawking}. Interestingly, the phase of the time-periodic modulation has also been shown to create an effective gauge potential and magnetic field for photons, which allows controlling photons as if they were charged particles under a real magnetic field, inducing Aharonov-Bohm-like effects and topological properties depending on the spatial distribution of the modulation phase \cite{fang2012photonic,fang2012realizing,fang2013controlling}. Finally, if different parts of the system are modulated with different phases in ways that break time-reversal symmetry, Lorentz reciprocity can also be broken \cite{sounas2017non} (in other words, by ``biasing'' the system with a physical quantity that is odd symmetric under time-reversal), with many intriguing implications as discussed in the next Section.

This brief overview highlights some of the most prominent and well-studied physical aspects of classical time-periodic (Floquet) systems. For additional details, we refer the interested readers to the many review articles that have been written about this burgeoning field, e.g., \cite{galiffi2022photonics,sounas2017non,yuan2018synthetic,kord2020microwave,caloz2020spacetime,lustig2023photonic,nagulu2020nonreciprocal,asgari2024theory}. In the following, we instead focus on two specific areas within this broad context that, in our opinion, deserve particular attention and may have a large impact in the near future. 

\section{Nonreciprocity}

Nonreciprocity has been one of the most successful applications of time modulation \cite{sounas2017non}. Nonreciprocal devices, such as isolators and circulators, are essential for protecting sources from reflections or enabling full-duplex operation. Furthermore, nonreciprocity has found applications in new areas in physics, such as topological photonics \cite{fleury2016floquet}. Time modulation allows the design of nonreciprocal components without magnetic materials, making their integration on chip easier. As mentioned above, in time modulated devices, nonreciprocity is achieved by modulating different parts of a device with different phases in ways that break time-reversal symmetry. For example, a common type of modulation is one with the form of a traveling wave (an early example of which being traveling-wave amplifiers \cite{cullen1958atravelling}). Applying such modulation to a waveguide and matching the phase velocities of the modulation signal and the waveguide mode, one can maximize the effect of modulation on the waveguide mode for one propagation direction and minimize it for the opposite one \cite{yu2009complete}. The asymmetry in wave transmission can manifest in different ways, including asymmetric transmission phase or mode conversion. Furthermore, by applying this type of modulation to resonant structures, such as ring resonators, one can reduce size and realize compact nonreciprocal devices.

A major challenge in the design of time-modulated nonreciprocal devices is the generation of modulation signals with different phases over different parts of the device. This difficulty is compounded by the fact that important device metrics, such as isolation, insertions loss, and bandwidth, are proportional to modulation index and the fractional modulation frequency (i.e., the ratio between the modulation and operation frequencies). This problem is especially prominent at optical frequencies, where refractive index modulation is usually small, and even small fractional modulation frequencies translate to significantly high modulation frequencies. One approach to resolve this issue is by increasing the interaction volume between the modulation and optical signals either physically or effectively through resonances \cite{dostart2021optical}. However, these strategies lead to large structures or small bandwidths. Overcoming these limitations requires the development of new materials and/or modulation mechanisms that provide greater modulation indices and modulation frequencies. Past years have witnessed promising advancements towards this end. For example, one promising direction is to use integrated thin-film lithium niobate, a platform that has recently experienced significant advancements \cite{yu2023integrated}. 
Another interesting approach is based on acousto-optical effects \cite{tian2021magnetic,sohn2021electrically,kittlaus2021electrically,wanjura2023quadrature}. While these effects typically require the use of resonant structures, thus leading to limited bandwidths, they exhibit low noise, making them suitable for applications in quantum optics. All-optical nonlinearities can also be used to realize nonreciprocity through frequency mixing \cite{abdelsalam2020linear} or refractive-index self-modulation \cite{yang2020inverse}. Although these approaches require strong signals, they are characteristic for their simplicity, and they could be attractive in high-power applications. All these approaches are expected to further advance in the following years and lead to novel designs of nonreciprocal devices across the optical spectrum.

Another area where one can expect significant advances is the design of nonreciprocal metasurfaces or other metamaterial structures. Nonreciprocal metasurfaces could be useful for isolation in free-space optics and for nonreciprocal beam steering. Furthermore, they could be used to improve the performance of photovoltaic and thermal emission systems \cite{park2021reaching}. Although several works have theoretically analyzed time-modulated metasurfaces \cite{hadad2015space,wang2020theory,wu2020space,barati2020topological}, experimental studies have been limited \cite{guo2019nonreciprocal,shaltout2019spatiotemporal}, due to the difficulties in realizing spatio-temporal modulations over extended areas, as required in metasurfaces. However, one can expect that the advances in guided-wave time-modulated systems mentioned earlier can also solve some of the implementation problems of time-modulated metasurfaces. At microwave frequencies, the design of spatiotemporally modulated metasurfaces could be assisted by recent advances in phased antenna arrays, which decentralize the generation of radiated signals through arrays of oscillators overlapping the antenna array \cite{sengupta20120}. A similar approach could be applied to metasurfaces for the generation of the modulation signals through arrays of phase-coupled oscillators overlapping the metasurface elements \cite{kadry2020angular}, overcoming the challenges related to distributing the signal of a central source to various parts of the structure. Another approach could be based on the generation of an effective modulation in nonlinear materials through waves that break time-reversal symmetry, such as circularly polarized beams \cite{duggan2019optically,duggan2020nonreciprocal}. This strategy could be applied to both microwave and optical systems, opening opportunities for the implementation of nonreciprocal metasurfaces across the entire electromagnetic spectrum. Electro-optic modulation is another possible direction for the design of spatiotemporally modulated metasurfaces, assisted by the recent progress in the nano-patterning of materials with strong electro-optical properties, such as lithium niobate, multi-quantum wells \cite{mann2021ultrafast}, or two-dimensional polaritonic materials \cite{yu20172d}. The high modulation indices of these materials could overcome the need for strong resonances. 
All these approaches hold great promise for the realization of spatiotemporally modulated metasurfaces and are expected to lead to significant advancements in the following years.

\section{Breaking theoretical limitations and performance bounds}
\begin{figure}
    \centering
    \includegraphics[width=1.1\linewidth]{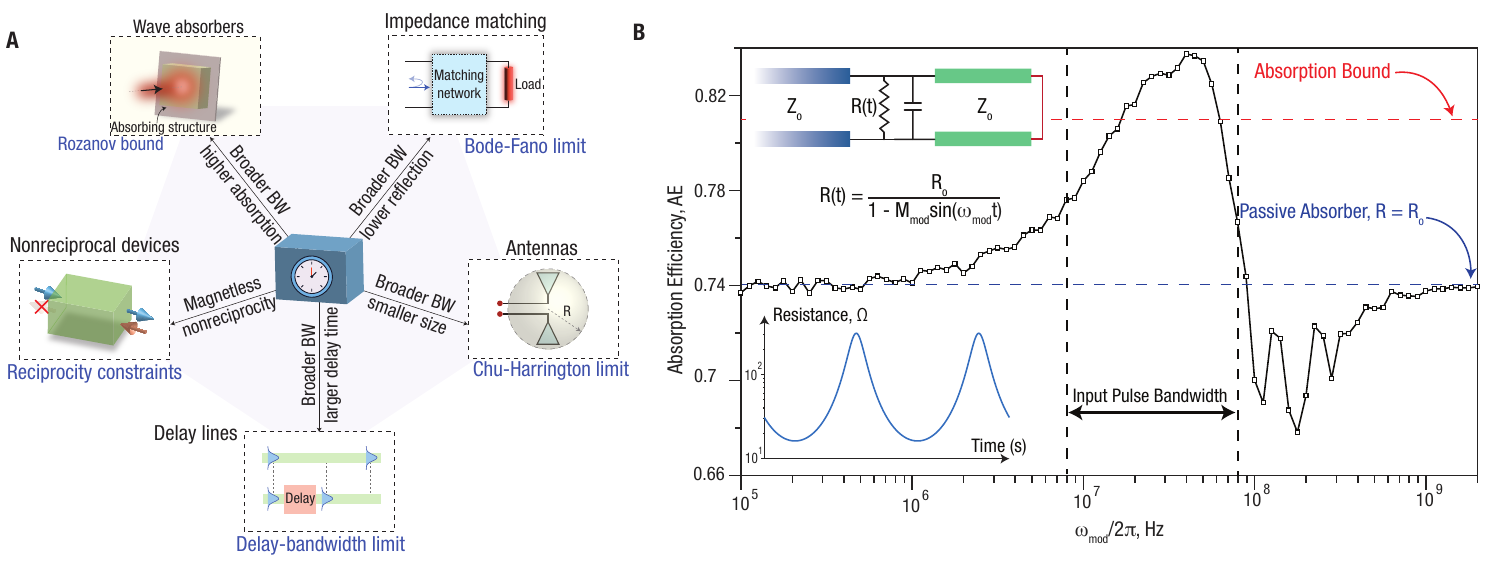}
    \caption{(A) Examples of limits, constraints, and application domains that can be impacted by time-varying systems. Adapted from \cite{hayran2024timevarying}. (B) Time-periodic absorber operating beyond the limits of linear time-invariant systems. Simulation results of the absorption efficiency of an electromagnetic absorber based on a time-periodic resistive elements inducing destructive interference between harmonics, as in \cite{ciabattoni2024observation}.
    The inset shows the equivalent circuit of the absorber and the temporal variation of the resistance.}
    \label{fig1}
\end{figure}

Another major aspect of time-periodic modulations that we believe deserves particular attention is their potential to improve the performance of various classes of devices in terms of operational bandwidth, size, and other metrics Fig. \ref{fig1}(a). Such improvements may even exceed what is normally allowed by well-established theoretical limits that have been identified in different areas of wave physics and engineering. While the idea of harnessing temporal degrees of freedom for this goal has started to attract attention \cite{hayran2023using,asgari2024theory}, this research area is still in its infancy.

Some of the theoretical limits and performance bounds that can potentially be broken through the application of temporal modulations include: (i) The \emph{Chu-Harrington limit}, which determines the maximum possible bandwidth of an electrically small antenna (or any radiating system) as a function of its maximum dimension \cite{chu1948physical}; (ii) the \emph{Bode-Fano limit on broadband impedance matching}, which expresses a fundamental trade-off between achievable reflection reduction and bandwidth for any lossless impedance-matching network applied to a given load \cite{fano1950theoretical}; (iii) the \emph{Rozanov bound on electromagnetic absorption} \cite{rozanov2000ultimate}, also known as the ``causality limit'' in the acoustics literature \cite{yang2018anintegration}, which is a limit on the maximum possible bandwidth of an absorber as a function of its thickness; (iv) \emph{theoretical limits on the delay-bandwidth product}, which imply that one cannot arbitrarily delay a signal independently of its bandwidth, with the delay-bandwidth product limited by some general properties of the system, e.g., its length \cite{miller2007fundamental}. These, and other, physical bounds and theoretical limits are derived under a combination of several assumptions, usually causality, passivity, linearity, and time-invariance; hence, they are not truly ``fundamental,'' as all these assumptions except causality can in principle be broken, opening intriguing opportunities to go well beyond what is possible with conventional linear time-invariant (LTI) passive designs. In particular, if the assumptions of time-invariance and/or linearity are broken, causality still holds, but, in most cases (though not all) it no longer makes sense to define a single-frequency transfer function, as frequency is not conserved. The derivation of Kramers-Kronig-like relations, sum rules, and related bounds, such as the Bode-Fano limit and the Rozanov bound, that depend on the analytic and asymptotic properties of a single-frequency transfer function must therefore be carefully re-evaluated \cite{solis2021functional,koutserimpas2024timevarying,lucarini2005kramerskronig}. 
Moreover, as mentioned in the Introduction, passivity may also be violated due to the interaction between signal(s) and modulation/pump.
 
Many of the features of linear time-periodic systems, and how they may impact relevant performance bounds, can also be understood from a circuit perspective. In a linear circuit (which could be an actual electrical circuit or a circuit model of a given structure), one can modulate either reactive or resistive components. If a reactive component is modulated, for example a capacitance (corresponding to modulating the real part of the permittivity of a material), the time-domain current-voltage relation acquires an additional term: $I(t) = \frac{dC(t)V(t)}{dt} = C(t)\frac{dV(t)}{dt} + V(t)\frac{dC(t)}{dt}$, with the last term leading to energy transfer with the modulation. This can be seen by calculating the total energy dissipated, or gained, by the system: $E=\int_{-\infty}^{\infty} V(t) I(t) dt$, which can be calculated to be equal to $\frac{1}{2} \int_{-\infty}^{\infty} V(t)^2 (dC(t)/dt) dt$ for any signal of finite temporal width, implying the possibility of non-zero power dissipated/gained if the capacitance is time-varying. Note that the possibility of transferring energy to/from the modulation does not necessarily imply the presence of gain and it is still an open question under what general conditions one might expect that a given performance bound is broken. One promising direction is the following: If parametric gain is present, this may not only amplify propagating signals (which by itself may break certain efficiency tradeoffs of passive systems), but may also lead to anomalous dispersion (``non-Foster dispersion'' in the applied electromagnetic literature \cite{sussman2009nonfoster,jacob2016gain}) due to the conversion of positive to negative frequencies associated with this gain process and the resulting inversion of the frequency dispersion of the impedance (or any other constitutive parameter). This fact could then be used to break theoretical limitations, especially bandwidth limits such as the Bode-Fano limit, that assume (implicitly or explicitly) a normal dispersion response (monotonically increasing with frequency in low-loss regions). Different from the case of a time-varying reactance, if a resistive component is modulated (corresponding to modulating the conductivity of a material), the current-voltage relation is Ohm’s Law, $V(t)=R(t)I(t)$, which has the same form for time-varying and time-invariant resistors, with no additional terms, implying that no energy needs to be exchanged with the modulator \cite{mostafa2022coherently,ciabattoni2024observation}. As a result, one can generate frequency harmonics and change the frequency content of a signal without adding external energy. The generation of harmonics and their possible interference can then alter the absorption of energy by the time-varying resistor. This effect could then allow breaking performance limitations related to wave absorption, energy harvesting, and reflection/scattering reduction through dissipation.

Along these lines, as an example illustrating some of the possibilities enabled by time-periodic modulations, two general mechanisms have been recently studied to increase absorption beyond the bandwidth-thickness limits predicted by the ``Rozanov bound'' mentioned above: (i) A possible strategy is to use ``parametric absorption,'' the time-reversed equivalent of parametric gain, in a lossy time-periodic slab, increasing absorption beyond the limits of its time-invariant counterpart by transferring energy from the propagating wave to the modulating agent \cite{hayran2024beyond}; (ii) a second approach is to rely instead on the periodic modulation of a purely resistive element to induce destructive interference between reflected harmonics, hence reducing total reflection and dissipating more energy in the time-varying resistor, over a broad bandwidth \cite{mostafa2022coherently,ciabattoni2024observation}. As shown in Fig. \ref{fig1}, for a broadband pulse incident on a sinusoidally time-modulated resistance above a ground plane, the absorption efficiency, defined as in \cite{ciabattoni2024observation}, rises above the absorption limit of LTI systems as the modulation frequency $\omega_{\text{mod}}$ increases. Absorption then drops for $ \omega_{\text{mod}} \gtrapprox 60$ MHz, which can be understood from the more limited interaction/interference between harmonics as the modulation-induced frequency transitions start exceeding the bandwidth of the incident signal. This strategy, which was recently experimentally demonstrated \cite{ciabattoni2024observation}, shows that it is possible to break LTI absorption bounds even with a perfectly passive linear time-periodic system, and with no energy exchange with the modulation. Clearly, however, any strategy based on interference between harmonics only works for temporally coherent incident signals, making this approach more practically relevant for applications requiring scattering reduction under pulsed excitation, as in radar stealth technology.

More broadly, while the possibility of using time-periodic systems to break LTI performance limits is exciting, several open questions remain. Many of the approaches described above would need to be tuned/optimized depending on some properties of the incoming signal. This calls for the development of new and improved strategies to address this issue, for example using phase-insensitive processes or self-configuring approaches. From the theoretical standpoint, one of the most important goals for future work is, in our opinion, the development of a more systematic understanding of what specific features of time-periodic systems allow breaking various LTI limits and under what conditions, as well as the possible extension of these theoretical limits to time-varying scenarios. Are there more fundamental limits to classical wave radiation, impedance matching, absorption, etc. that apply to at least some classes of time-varying systems? Can general considerations related to stability, noise figures, and information capacity help define the ultimate performance bounds of these active systems? More research is certainly needed to address these and other questions at the foundation of this research area. 

\begin{backmatter}
\bmsection{Funding}
Air Force Office of Scientific Research with Grant No. FA9550-22- 1-0204; Office of Naval Research with Grant No. N00014-22-1-2486.

\bmsection{Disclosures}
The authors declare no conflicts of interest.

\bmsection{Data Availability Statement}
The data used to produce these results is available upon reasonable request to the corresponding author.

\end{backmatter}

\bibliography{bibliography}

\end{document}